\documentclass[a4paper]{jpconf}
\usepackage{graphicx}
\usepackage{amsmath,amssymb,latexsym}
\usepackage{xcolor}
\usepackage{comment}

\begin{document}
\title{A novel approach to autoparallels for the theories of symmetric teleparallel gravity}

\author{Caglar Pala$^1$, Muzaffer Adak$^2$}

\address{$^1$ Erciyes University, 38030 Kayseri, Turkey}
\address{$^2$ Pamukkale University, 20017 Denizli, Turkey}

\ead{$^1$caglar.pala@gmail.com, $^2$madak@pau.edu.tr}

\begin{abstract}
Although the autoparallel curves and the geodesics coincide in the Riemannian geometry in which only the curvature is nonzero among the nonmetricity, the torsion and the curvature, they define different curves in the non-Riemannian ones. We give a novel approach to autoparallel curves and geodesics for theories of the symmetric teleparallel gravity written  in the coincident gauge. Then we apply our autoparallel equation to a Schwarzschild-type metric and give remarks about dark matter and orbit equation.\\


{\noindent \it \normalsize Dedicated to Tekin Dereli in honour of his 72$^{nd}$ birthday}
\end{abstract}

\section{Introduction}
There have been two main ingredients for the theories of gravity since Isaac Newton; field equations and trajectory equations. The field equations are cast for describing the dynamics of a source and the trajectory ones are the for determining path followed by a spinless massive point test particle. For the Newton's theory of gravity they are written respectively as
 \begin{subequations}
 \begin{align}
     \nabla^2 \phi &= 4\pi G \rho \, ,   \\
     \frac{d^2 \vec{r}}{dt^2} + \vec{\nabla} \phi &=0  \, ,
 \end{align}
 \end{subequations}
where $G$ is the Newton's constant of gravity,  $\vec{r}$ is the three-dimensional Euclidean position vector of the test particle, $t$ is the time disjoint from the Euclidean space, $\rho$ is the volume density of mass of the source, $\vec{\nabla}$ is the gradient operator in the three-dimensional Euclidean space and $\phi$ is the gravitational field from a source at the position of $\vec{r}$. On the other hand, in the Einstein theory of gravity, the general theory of relativity, the effect of the spacetimes curvature on matter fields is interpreted as the gravitational force and one assumes that a massive point test particle has spacetime history that matches to the time-like geodesic associated with the spacetime metric.
 \begin{subequations}
 \begin{align}
     R_{\mu\nu}-\frac{1}{2}Rg_{\mu \nu}&=4\pi G T_{\mu\nu} \, , \\
     \frac{d^2 x^\mu}{d\tau^2} + \widetilde{\omega}^\mu_{\nu \lambda}  \frac{dx^\nu}{d\tau}\frac{dx^\lambda}{d\tau}&=0 \, ,\label{eq:geodesic1}
 \end{align}
 \end{subequations}
where $g_{\mu\nu}$ is the metric tensor of $(1+3)$-dimensional pseudo-Riemannian spacetime, $R_{\mu\nu}$ is its Ricci curvature tensor, $R$ is its curvature scalar, $T_{\mu \nu}$ is the energy-momentum tensor for the source, $\tau$ is a affine parameter (proper time in practice), $x^\mu$ denotes a chosen coordinate system of the spacetime, $ \widetilde{\omega}^\mu_{\nu \lambda} $ are the Christoffel symbols (the Levi-Civita connections) of the metric. The first equation is known as the Einstein's equation and the second one is geodesic equation of the pseudo-Riemannian spacetime. 

These prescriptions were compatible with the observations for a while. But there are pretty much reasons to go beyond Einstein's theory of
gravity coming from the experimental side such as dark matter and dark energy and also from theoretical side such as vain efforts of obtaining a quantum gravity, see for example \cite{fhehl1995}-\cite{madak2010}. The correlation between the general relativity and the geometry of
the spacetime makes it natural to explore modifications to the general relativity based on extending its geometrical framework. In these cases, the geodesic hypothesis based on autoparallels of the Levi-Civita connection may need to be extended in alternative theories of gravity formulated in the non-Riemannian geometries. Shortly the geodesic equation (\ref{eq:geodesic1}) is replaced with the autoparallel equation
  \begin{equation}
  \frac{d^2 x^\mu}{d\tau^2} + \omega^\mu_{\nu \lambda}  \frac{dx^\nu}{d\tau}\frac{dx^\lambda}{d\tau}=0 \label{eq:autoparallel1}
 \end{equation}
where $\omega^\mu_{\nu \lambda}$ are the non-Riemannian connections containing torsion and/or nonmetricity apart from the Levi-Civita connections. For example, in the reference \cite{tdereli2001} the authors investigated the geodesic hypothesis based on autoparallels of the Levi-Civita connection in the Brans-Dicke theory in the Riemann-Cartan spacetime and thereby found that if the Brans-Dicke scalar field couples significantly to matter and test particle moves on such worldline, the commenly accepted methods based on the conventional geodesic hypothesis may need refinement. Those results were improved in the work \cite{cebeci2004} in which it was of interest to compare the autoparallel orbits derived from different metric compatible spacetime connections for a spinless test particle in the metric of a spinning source. In a recent work \cite{obukhov2021}, autoparallels in alternative gravity have been treated again. The authors explicitly derived autoparallels as effective post-Riemannian geometric constructs, and at the same time they argued against postulating autoparallels as fundamental equations of motion for test bodies in alternative gravity theories.

There is not an agreement on which equation, (geodesic of the Levi-Civita connection or autoparallel of the full connection) should be used as the worldline of a spinless test particle at vicinity of a gravitating source in non-Riemennian spacetime geometries. Experiments will guide us whether particles follow metric geodesic paths or their trajectories are in turn autoparallel curves for the full affine connection \cite{mao2007prd}. However, we can discuss which one seems more natural from a theoretical perspective. At this context the reference \cite{jimenez2020} states that metric geodesic trajectories seem better aligned with our current understanding of physics.

One of the alternative methods for modifying Einstein's theory of general relativity written in the Riemannian geometry is to propose new theories in a non-Riemannian geometry. Together with our collaborators we have published a series of researches on the modified gravity theories formulated in terms of only nonmetricity
without torsion and curvature, the so-called symmetric teleparallel gravity \cite{madak2005}-\cite{madak2018} just after the pioneering work \cite{jmnester1999}. These works have attracted noteworthy interest in the scientific community \cite{imol2017}-\cite{hohmann2021}. All the works above investigated new gravity models by starting from a Lagrangian. Then the field equations have been obtained via variation. Finally some possible solutions have been obtained and analyzed. But the trajectory of a test particle has not been discussed so much in the theories of symmetric teleparallel gravity \cite{miranda2019},\cite{damianos2019}. Therefore this complementary work was performed. In a very recent work in this direction \cite{yang2021}, the authors consider the geodesic deviation equation, describing the relative accelerations of nearby particles, and the Raychaudhuri equation, giving the evolution of the kinematical quantities associated with deformations in the Weyl-type $f(Q,T)$ gravity, in which the nonmetricity $Q$ is represented in the standard Weyl form, fully determined by the Weyl vector, while $T$ represents the trace of the matter energy–momentum tensor.

\section{Mathematical preliminaries}

Spacetime, in general, is denoted by $\{ M,g,\nabla\}$ where $M$
is orientable and differentiable manifold, $g=g_{\alpha \beta}
e^\alpha \otimes e^\beta$ is the metric tensor written in terms of 1-forms (or coframe) $e^\alpha$ and $\nabla$ is connection associated with connection 1-forms ${\omega^\alpha}_\beta$. In the language of the exterior algebra, the Cartan structure equations
define nonmetricity 1-forms, torsion 2-forms and curvature
2-forms, respectively
 \begin{subequations}\label{cartaneqns}
 \begin{align}
     Q_{\alpha \beta} &:= - \frac{1}{2} D g_{\alpha \beta}
                      = \frac{1}{2} (-d g_{\alpha \beta} + \omega_{\alpha \beta}+\omega_{\alpha \beta}) \, , \label{nonmet}\\
     T^\alpha &:= D e^\alpha = d e^\alpha + {\omega^\alpha}_\beta \wedge e^\beta \, , \label{tors}\\
     {R^\alpha}_\beta &:= D {\omega^\alpha}_\beta := d {\omega^\alpha}_\beta
                  + {\omega^\alpha}_\gamma \wedge {\omega^\gamma}_\beta \, , \label{curva}
 \end{align}
 \end{subequations}
where $\wedge$ denotes the exterior product, $d$ the exterior derivative and $D$ the covariant exterior derivative \cite{ecartan1923}. Geometrically, the nonmetricity tensor measures the deformation of length and angle standards during parallel transport. For example, after parallel transportation of a vector along a closed curve, the length of the final vector may be different from that of the initial vector. On the other hand, torsion relates to the translational group. Moreover it is sometimes said that closed parallelograms do not exist in spacetime with torsion. It can be restated in terms of
the picture of manifold and bundle as that after a vector is
parallel carried around a closed path on a manifold, if torsion is not vanishing, there will be an interval between initial and final points on the fiber of the bundle over the manifold. Finally, curvature is related to the linear group. That is, when a vector is parallel transported along a closed loop, the vector undergoes a rotation due to curvature. These three tensors satisfy the Bianchi identities:
 \begin{subequations}
 \begin{align}
       D Q_{\alpha \beta} &= \frac{1}{2} ( R_{\alpha \beta} +R_{\beta \alpha}) \, , \label{bianc:0} \\
       D T^\alpha    &= {R^\alpha}_\beta \wedge e^\beta \, , \label{bianc:1} \\
       D {R^\alpha}_\beta &= 0  \, . \label{bianc:2}
 \end{align}
 \end{subequations}
The full connection 1-forms can be decomposed uniquely as follows \cite{fhehl1995}-\cite{madak2010}, \cite{pala2021}:
 \begin{equation}
     {\omega^\alpha}_\beta =  \underbrace{(g^{\alpha \gamma} d g_{\gamma_\beta} + {p^\alpha}_\beta)/2 + {\widetilde{\omega}^\alpha}{}_\beta}_{Metric}
     + \underbrace{{K^\alpha}_\beta}_{Torsion} + \underbrace{{q^\alpha}_\beta  + {Q^\alpha}_\beta}_{Nonmetricity}   \label{connect:dec}
  \end{equation}
where $\widetilde{\omega}^\alpha{}_\beta$ Levi-Civita connection 1-forms
 \begin{equation}
     \widetilde{\omega}^\alpha{}_\beta \wedge e^\beta = -d e^\alpha  \; , \label{LevCiv}
 \end{equation}
${K^\alpha}_\beta$ contortion tensor 1-forms,
 \begin{equation}
   {K^\alpha}_\beta  \wedge e^\beta = T^\alpha  \; , \label{contort}
 \end{equation}
and anti-symmetric 1-forms
 \begin{align}
   &q_{\alpha \beta} = -( \iota_\alpha  Q_{\beta \gamma } ) e^\gamma + ( \iota_\beta Q_{\alpha \gamma})
    e^\gamma  \; , \label{q:ab} \\
   &p_{\alpha \beta} = -( \iota_\alpha  d g_{\beta \gamma } ) e^\gamma + ( \iota_\beta d g_{\alpha \gamma})
    e^\gamma \label{p:ab}\; ,
  \end{align}
where $\iota_\alpha$ denotes the interior product with respect to the basis vector $\partial_\alpha$ dual to $e^\alpha$; $e^\alpha (\partial_\beta)= \delta^\alpha_\beta$. This decomposition is self-consistent. To see that it is enough to
multiply (\ref{connect:dec}) from right by $e^\beta$ and to use
definitions above. As moving indices vertically in front of both $d$ and $D$, a special attention is needed because in general it may be $d
g_{\alpha \beta} \neq 0$ and $D g_{\alpha \beta} \neq 0$. Symmetric part of the full connection comes from (\ref{nonmet})
 \begin{equation}
  \omega_{(\alpha \beta)} = Q_{\alpha \beta } + \frac{1}{2} d g_{\alpha \beta } \label{connect:sym}
  \end{equation}
and the remainder is anti-symmetric part
 \begin{equation}
  \omega_{[\alpha \beta]} = \frac{1}{2} p_{\alpha \beta} + \widetilde{\omega}_{\alpha \beta} + K_{\alpha \beta} + q_{\alpha \beta}  \; . \label{connect:ansym}
 \end{equation}
In a holonomic or coordinate basis in which $e^\alpha = dx^\alpha$, then $de^\alpha=0$ via the Poincare lemma, this decomposition becomes
  \begin{equation}\label{decomp-con-coord}
{\omega^\alpha}_\beta = \frac{1}{2} g^{\alpha \sigma}(\partial_\gamma g_{\sigma \beta} + \partial_\beta g_{\sigma \gamma} - \partial_\sigma  g_{\beta \gamma})e^\gamma + {K^\alpha}_\beta + {q^\alpha}_\beta  + {Q^\alpha}_\beta
 \end{equation}
where we used $\iota_\alpha d \equiv \partial_\alpha$. We notice that the first group on the right hand side coincide with the Christoffel symbols. From now on we always work with the coordinate frame.

\section{The coincident gauge in the symmetric teleparallel geometry}

If only $Q_{\alpha \beta}=0$, the connection is said to be metric compatible; Einstein-Cartan geometry. If both $Q_{\alpha \beta}=0$ and $T^\alpha =0$, the connection is called Levi-Civita; pseudo-Riemannian geometry. If both $R^\alpha{}_\beta=0$ and $Q_{\alpha \beta}=0$, it is called teleparallel (Weitzenb\"{o}ck) geometry. If both $R^\alpha{}_\beta=0$ and $T^\alpha =0$, it is called symmetric teleparallel (or Minkowski-Weyl) geometry \cite{mao2007prd}. That is, in the symmetric teleparallel geometry only the nonmetricity tensor is nonzero: $Q_{\alpha \beta} \neq 0 \, , \quad T^\alpha =0 \, , \quad {R^\alpha}_\beta  =0$. Thus the Cartan structure equations (\ref{cartaneqns}) give a set of nonlinear dynamical equation for the full connection, ${\omega^\alpha}_\beta$, in coordinate (holonomic) frame in which $e^\alpha = dx^\alpha$ and then $de^\alpha=0$,
 \begin{subequations}\label{STPG1}
 \begin{align}
     \frac{1}{2} (-d g_{\alpha \beta} + \omega_{\alpha \beta}+\omega_{\alpha \beta}) &\neq 0 \; , \label{nonmet1}\\
      {\omega^\alpha}_\beta \wedge dx^\beta &=0 \; , \label{tors1}\\
      d {\omega^\alpha}_\beta + {\omega^\alpha}_\gamma \wedge {\omega^\gamma}_\beta &=0 \, . \label{curva1}
 \end{align}
 \end{subequations}
Here the full connection 1-form can not be solved analytically. Meanwhile, it is worthy to remember that we could have done it in the Riemannian geometry. However, one special solution is to choose ${\omega^\alpha}_\beta = 0$ yielding $Q_{\alpha \beta} = - \frac{1}{2} d g_{\alpha \beta} \neq 0 \, , \quad T^\alpha =0 \, , \quad  {R^\alpha}_\beta =0$. This choice is called the {\it coincident gauge}\footnote{When we firstly realized this trick, we called it as {\it a gauge fixing} in our paper \cite{madak2013}, later we read the nomenclature {\it coincident gauge} in the literature.}.

 \subsection{Autoparallels}

In the Riemannian geometry one requires that the tangent vector, $\mathcal{T}^\mu$, to the autoparallel curve, $x^\mu(\tau)$, points in the same direction as itself when parallel propagated, and demands that it maintains the same length
\cite{rwald1984}
 \begin{equation}
    D \mathcal{T}^\mu =0 \; .
 \end{equation}
On the other hand, since intuitively autoparallel curves are
``those as straight as possible" we do not demand the vector to
keep the same length during parallel propagation in STPG. It is
known that nonmetricity is related to length and angle standards
at parallel transportation. Therefore we prescribe the parallel
propagation of the tangent vector as
 \begin{equation}
     D \mathcal{T}^\mu = (a Q^\mu{}_\nu +bq^\mu{}_\nu ) \mathcal{T}^\nu + c Q \mathcal{T}^\mu
     \label{eq:DTmu}
 \end{equation}
where $\mathcal{T}^\mu = {d x^\mu(\tau)}/{d \tau}$ is the tangent vector to the curve $x^\mu (\tau)$ with an affine parameter $\tau$, $Q^\mu{}_\nu$ is the nonmetricity 1-form, $Q=Q^\mu{}_\mu$ is the Weyl 1-form, $q^\mu{}_\nu$ is defined by (\ref{q:ab}) in terms of the nonmetricity and $a,b,c$ are arbitrary constants. This prescription of parallel transport is our novel approach. Here since we are in the coincident gauge, $\omega^\alpha{}_\beta =0$, we obtain $D \mathcal{T}^\mu = d\mathcal{T}^\mu = (\partial_\alpha \mathcal{T}^\mu) dx^\alpha$. Moreover, we write all one-forms in their components $Q^\mu{}_\nu=Q^\mu{}_{\nu \alpha} dx^\alpha$, $q^\mu{}_\nu = q^\mu{}_{\nu\alpha} dx^\alpha$ and $Q=
Q^\nu{}_{\nu\alpha} dx^\alpha$. Then the equation (\ref{eq:DTmu}) yields
 \begin{equation}
  \partial_\alpha \mathcal{T}^\mu =(a Q^\mu{}_{\nu \alpha} + b
  q^\mu{}_{\nu \alpha}) \mathcal{T}^\nu + c Q^\nu{}_{\nu \alpha} \mathcal{T}^\mu \, .
  \end{equation}
Now after multiplying  this with $\mathcal{T}^\alpha$ and then writing $\mathcal{T}^\alpha \partial_\alpha := \frac{d}{d\tau}$ on the left hand side and $\mathcal{T}^\mu = {d x^\mu}/{d \tau}$ on the both sides, we arrive at
 \begin{equation}
  \frac{d^2 x^\mu}{d\tau^2} = (a Q^\mu{}_{\nu \alpha} + b
  q^\mu{}_{\nu \alpha}) \frac{d x^\nu}{d \tau} \frac{d x^\alpha}{d \tau}   + c Q^\nu{}_{\nu \alpha} \frac{d x^\mu}{d \tau} \frac{d x^\alpha}{d \tau} \, .
 \end{equation}
Here by using $Q_{\alpha \beta} = -\frac{1}{2} d g_{\alpha
\beta}$, the result of the coincident gauge, we get $Q^\mu{}_{\nu \alpha} = -\frac{1}{2}g^{\mu \beta}(\partial_\alpha g_{\beta \nu})$ and accordingly $q^\mu{}_{\nu \alpha} = -\frac{1}{2}g^{\mu \beta}(\partial_\nu g_{\alpha \beta} - \partial_\beta g_{\alpha \nu})$. Consequently, we express the autoparallel curve equation in terms of the metric as follows
 \begin{equation}
  \frac{d^2 x^\mu}{d\tau^2} = g^{\mu \beta} \left[ - \frac{a+b}{4} (\partial_\alpha g_{\beta \nu} + \partial_\nu g_{\beta \alpha})
  + \frac{b}{2} (\partial_\beta g_{\alpha \nu}) \right] \frac{d x^\nu}{d \tau} \frac{d x^\alpha}{d \tau} 
   - \frac{c}{2}g^{\nu \beta} (\partial_\alpha g_{\beta \nu}) \frac{d x^\mu}{d \tau} \frac{d x^\alpha}{d \tau} \label{eq:autoparalleleq2}
 \end{equation}
where we symmetrized the first term of the square parentheses $(\nu \alpha)= \frac{1}{2}(\nu \alpha + \alpha \nu)$. We notice that if we fix $a=b=1$ and $c=0$, then this equation becomes the same as the autoparallel curve of the Riemannian geometry.

\subsection{Geodesics}

Intuitively, geodesics are ``curves as short as possible".
An interval between two infinitesimal points is given by the metric
 \begin{equation}
    ds^2 = g_{\mu \nu} dx^\mu dx^\nu
 \end{equation}
we can parameterize the curve between endpoints as $x^\mu =
x^\mu(\tau)$, then we obtain (in a metric with signature $(-,+,+,+)$)
 \begin{equation}
     s=\int_{\tau_1}^{\tau_2} \left( -g_{\mu \nu} \dot{x}^\mu \dot{x}^\nu \right)^{1/2}d\tau
 \end{equation}
where dot denotes a derivative with respect to the affine parameter, $\tau$. We inserted a minus sign because of the Lorentz signature. We wish now to derive the condition on a curve which makes it extremize the length between its endpoints, i.e., wish to find those curves whose length does not change to first order under an arbitrary smooth deformation which keeps the endpoints fixed. This condition gives rise to the Euler-Lagrange equations
 \begin{equation}
     \frac{d}{d \tau} \left( \frac{\partial L}{\partial \dot{x}^\alpha} \right) - \frac{\partial L}{\partial
     x^\alpha}=0
 \end{equation}
of the action integral $I=\int_{\tau_1}^{\tau_2} L(x^\mu ,
\dot{x}^\mu , \tau)$. Thus in our case the Lagrangian is
 \begin{equation}
    L = \left[ -g_{\mu \nu} (x) \dot{x}^\mu \dot{x}^\nu \right]^{1/2}
 \end{equation}
where $x$ stands for coordinate functions $x^\mu$. Now the Euler-Lagrange equations yield
 \begin{equation}
   \ddot{x}^\beta + \frac{1}{2} g^{\alpha \beta} (\partial_\mu g_{\alpha \nu} + \partial_\nu g_{\alpha \mu}
   + \partial_\alpha g_{\mu \nu})\dot{x}^\mu \dot{x}^\nu =
   \frac{\dot{x}^\beta}{2 ( g_{\mu \nu} \dot{x}^\mu \dot{x}^\nu )}   \frac{d(g_{\mu \nu} \dot{x}^\mu
   \dot{x}^\nu)}{d\tau} \label{eq:geodesic}
 \end{equation}
Here we pay a special attention to the last term since the length of a vector needs not to be stay the same during the parallel transportation because of the existence of the nonmetricity. Let us evaluate it. First, we write it as $ g_{\mu \nu} \dot{x}^\mu \dot{x}^\nu =
g_{\mu \nu} \mathcal{T}^\mu \mathcal{T}^\nu = \mathcal{T}_\mu \mathcal{T}^\mu$. Now,
 \begin{align}
    d (\mathcal{T}_\mu \mathcal{T}^\mu) = (D g_{\mu \nu}) \mathcal{T}^\mu \mathcal{T}^\nu +2g_{\mu \nu} \mathcal{T}^\mu (D \mathcal{T}^\nu) = -2Q_{\mu \nu} \mathcal{T}^\mu \mathcal{T}^\nu + 2g_{\mu \nu}\mathcal{T}^\mu (D \mathcal{T}^\nu)
 \end{align}
Here the usage of the equation (\ref{eq:DTmu}) gives
  \begin{equation}
    d ( \mathcal{T}_\mu \mathcal{T}^\mu) = 2(a-1)Q_{\mu \nu} \mathcal{T}^\mu \mathcal{T}^\nu + 2c Q \mathcal{T}_\mu \mathcal{T}^\mu
  \end{equation}
This means that if we choose $a=1$ and $c=0$, the geodesic equation of STPG (\ref{eq:geodesic}) turns out to be the same as the geodesic equation of the Riemannian geometry.

\section{An application to  a stationary, spherically axisymmetric  metric}

Geodesics of (time-like and null) of Schwarzschild geometry are in
very good agreement with solar experiments. Furthermore, shadow of the
M87* super massive black hole is also in very good agreement with Kerr null geodesics \cite{wielgus2020}. On the other hand, the observed baryonic matter in extragalactic systems does not match gravitational potential of general relativity. This discrepancy is called as dark matter. A classic example is that the rotation curves of disk galaxies become approximately flat $(v \approx const)$ when they should be decreasing in a Keplerian way $(v \propto r^{-1/2})$. Accordingly we try to see possible novel effects of our results concretely by dealing with some astrophysical events. Thus we consider a stationary, spherically axisymmetric metric which is written approximately (up to the first order) in the spherical coordinates $x^\mu=(t,r,\theta,\phi)$ and in natural gravitational units $(c=G=1)$ \cite{mao2007prd}
 \begin{equation}
     g = -\left( 1+  \frac{\mathcal{H} m}{r} \right)dt^2 + \left( 1+  \frac{\mathcal{F} m}{r} \right)dr^2 + r^2 \left( d\theta^2 + \sin^2\theta d\phi^2 \right) +  \frac{2 \mathcal{G} m\ell}{r} \sin^2\theta dt d\phi
 \end{equation}
where $m$ and $J=m\ell$ are mass and rotational angular momentum of gravitating object, $\mathcal{H}, \mathcal{F}, \mathcal{G}$ are dimensionless constants. We remark that in general relativity, the Kerr metric at large distances yields these constants as $\mathcal{H}=- \mathcal{F}= \mathcal{G}=-2$. One can readily read the components
 \begin{align}
     g_{tt}&=-\left( 1+\frac{\mathcal{H}m}{r}\right), & g_{rr}&=1+\frac{\mathcal{F}m}{r}, & g_{\theta \theta} &=r^2, \nonumber \\
     g_{\phi \phi} &= r^2\sin^2\theta, & g_{t\phi}&= \frac{\mathcal{G}m\ell}{r}\sin^2\theta, & \text{other} \; g_{\alpha\beta}&=0,
 \end{align}
and compute the inverses up to the first order (practically by omitting the terms containing $\mathcal{H}^2, \mathcal{F}^2, \mathcal{G}^2, \mathcal{HG}, \mathcal{HF}, \mathcal{FG}$)
  \begin{align}
     g^{tt}&= -\left( 1-\frac{\mathcal{H}m}{r}\right), & g^{rr}&= 1-\frac{\mathcal{F}m}{r}, & g^{\theta \theta} &= \frac{1}{r^2}, \nonumber \\
     g^{\phi \phi} &=  \frac{1}{r^2\sin^2\theta}, & g^{t\phi}&=  \frac{\mathcal{G}m\ell}{r^3}, & \text{other} \; g^{\alpha\beta}&=0.
 \end{align}
Now we calculate the equation (\ref{eq:autoparalleleq2}) explicitly. With the notation $x^t := t$, $x^r := r$, $x^\theta := \theta$, $x^\phi := \phi$, for $\mu=\theta,t,\phi,r$, it reads respectively
 \begin{subequations}
  \begin{align}
     \frac{d^2\theta}{d\tau^2} =&  \left\{ -\frac{ a+b+2c }{r}  + \frac{c(\mathcal{H} + \mathcal{F})m}{2r^2}  \right\}  \frac{dr}{d\tau} \frac{d\theta}{d\tau} +  \frac{2b\mathcal{G}m\ell \sin\theta \cos\theta}{r^3}  \frac{dt}{d\tau} \frac{d\phi}{d\tau} \nonumber \\
    &+   b \sin\theta \cos\theta  \frac{d\phi}{d\tau} \frac{d\phi}{d\tau} -  c \cot\theta \frac{d\theta}{d\tau} \frac{d\theta}{d\tau} \, , \label{eq:thetaeqn} \\
     \frac{d^2t}{d\tau^2} =&  \left\{ -\frac{2c}{r} +   \frac{m\left[ (a+b)\mathcal{H} + c(\mathcal{H}+\mathcal{F}) \right]}{2r^2} \right\} \frac{dt}{d\tau} \frac{dr}{d\tau}
  -   c\cot\theta  \frac{dt}{d\tau} \frac{d\theta}{d\tau} \nonumber \\
  &- (a+b) \frac{3\mathcal{G}m\ell \sin^2\theta}{2r^2}  \frac{dr}{d\tau} \frac{d\phi}{d\tau}  \, , \label{eq:timeeqn}\\
     \frac{d^2\phi}{d\tau^2} =&   \frac{(a+b)\mathcal{G}m\ell}{2r^4} \frac{dt}{d\tau} \frac{dr}{d\tau} -  \frac{ (a+b)\mathcal{G}m\ell \cot\theta}{r^3} \frac{dt}{d\tau} \frac{d\theta}{d\tau} \nonumber \\
    &+  \left\{ -\frac{a+b+2c}{r} + \frac{c(\mathcal{H} + \mathcal{F})m}{2r^2}  \right\} \frac{dr}{d\tau} \frac{d\phi}{d\tau} -  (a+b+c)\cot\theta  \frac{d\theta}{d\tau} \frac{d\phi}{d\tau} \, , \label{eq:phieqn} \\
     \frac{d^2r}{d\tau^2} =&  \frac{b\mathcal{H}m}{2r^2} \frac{dt}{d\tau} \frac{dt}{d\tau} 
    +  \left\{ -\frac{2c}{r} + \frac{m\left[ c\mathcal{H} + (a+c)\mathcal{F} \right]}{2r^2}   \right\} \frac{dr}{d\tau} \frac{dr}{d\tau} -   c \cot\theta  \frac{dr}{d\tau} \frac{d\theta}{d\tau} \nonumber  \\
    &-  b\left( m \mathcal{F} - r \right) \frac{d\theta}{d\tau} \frac{d\theta}{d\tau}
    -  b\sin^2\theta \left( \mathcal{F}m  - r \right) \frac{d\phi}{d\tau} \frac{d\phi}{d\tau} -  \frac{b \mathcal{G}m\ell \sin^2\theta }{r^2}  \frac{dt}{d\tau} \frac{d\phi}{d\tau} \, . \label{eq:radialeqn}
 \end{align}
 \end{subequations}
Now we consider a body moving instantly at the equatorial plane. Formally we set $\theta=\pi/2$ and $d\theta/d\tau =0$ the equation (\ref{eq:thetaeqn}) and obtain $d^2\theta/d\tau^2=0$ which means that the body always stays at the equatorial plane. We substitute this result into (\ref{eq:timeeqn}), (\ref{eq:phieqn}) and (\ref{eq:radialeqn}),
 \begin{subequations} \label{eq:withrotation}
  \begin{align}
     \frac{d^2t}{d\tau^2} + \left\{ \frac{2c}{r} -   \frac{m\left[ (a+b)\mathcal{H} + c(\mathcal{H}+\mathcal{F}) \right]}{2r^2} \right\} \frac{dt}{d\tau} \frac{dr}{d\tau}
  + \frac{3(a+b)\mathcal{G}m\ell}{2r^2} \frac{d\phi}{d\tau} \frac{dr}{d\tau} &=0 \, , \label{eq:timeeqn2}\\
   \frac{d^2\phi}{d\tau^2} -   \frac{(a+b)\mathcal{G}m\ell}{2r^4}  \frac{dt}{d\tau} \frac{dr}{d\tau}
    + \left\{ \frac{a+b+2c}{r} - \frac{c(\mathcal{H} + \mathcal{F})m}{2r^2} \right\}  \frac{d\phi}{d\tau}\frac{dr}{d\tau} &=0 \, , \label{eq:phieqn2}\\
      \frac{d^2r}{d\tau^2} - \frac{b\mathcal{H}m}{2r^2}   \left(\frac{dt}{d\tau}\right)^2  
    +  \left\{ \frac{2c}{r} - \frac{m\left[ c\mathcal{H} + (a+c)\mathcal{F} \right]}{2r^2} \right\} \left(\frac{dr}{d\tau}\right)^2 & \nonumber \\
    +  b ( \mathcal{F}m  - r) \left(\frac{d\phi}{d\tau}\right)^2  + \frac{b\mathcal{G}m\ell }{r^2}  \frac{dt}{d\tau} \frac{d\phi}{d\tau} &=0 \, . \label{eq:radialeqn2}
   \end{align}
 \end{subequations}
In order to determine the constants of motion we rewrite the equation (\ref{eq:timeeqn2}) in terms of unknown functions $A(r)$ and $B(r)$
 \begin{align}
     \frac{d}{d\tau}\left[ A(r) \frac{dt}{d\tau} + B(r) \frac{d\phi}{d\tau} \right]=0
 \end{align}
which means that $A(r) \frac{dt}{d\tau} + B(r) \frac{d\phi}{d\tau}=const.$ We deliver the derivative and arrange the result.
 \begin{align}
     \frac{d^2t}{d\tau^2} + \frac{A'}{A} \frac{dr}{d\tau} \frac{dt}{d\tau} + \frac{B}{A} \frac{d^2\phi}{d\tau^2} + \frac{B'}{A} \frac{dr}{d\tau} \frac{d\phi}{d\tau}=0
 \end{align}
where prime denotes a derivative with respect to $r$. Now we insert the equation (\ref{eq:phieqn2}) into the third term.
 \begin{align}
      \frac{d^2t}{d\tau^2} + \left[ \frac{A'}{A} + \frac{B (a+b)\mathcal{G}m\ell}{2Ar^4} \right] \frac{dr}{d\tau} \frac{dt}{d\tau} + \left\{\frac{B'}{A} -\frac{B}{A} \left[ \frac{a+b+2c}{r} - \frac{cm(\mathcal{H} + \mathcal{F})}{2r^2}\right] \right\} \frac{dr}{d\tau} \frac{d\phi}{d\tau}=0
 \end{align}
By comparing this result with (\ref{eq:timeeqn2}), we arrive at two equations which can be used for determining the functions $A(r)$ and $B(r)$,
 \begin{subequations}
 \begin{align}
      \frac{A'}{A} + \frac{B (a+b)\mathcal{G}m\ell}{2Ar^4} &= \frac{2c}{r} -   \frac{m\left[ (a+b)\mathcal{H} + c(\mathcal{H}+\mathcal{F}) \right]}{2r^2} \, , \\
      \frac{B'}{A} -\frac{B}{A} \left[ \frac{a+b+2c}{r} - \frac{cm(\mathcal{H} + \mathcal{F})}{2r^2}\right] &= \frac{3(a+b)\mathcal{G}m\ell}{2r^2} \, .
 \end{align}
 \end{subequations}
Here since we obtain a very complicated set of solutions containing the Heun function via the software of Maple, for simplicity from this point we ignore the rotation by setting $\ell=0$. Then we rewrite the equations (\ref{eq:withrotation})
  \begin{subequations}
  \begin{align}
     \frac{d^2t}{d\tau^2} + \left\{ \frac{2c}{r} -   \frac{m\left[ (a+b)\mathcal{H} + c(\mathcal{H}+\mathcal{F}) \right]}{2r^2} \right\} \frac{dt}{d\tau} \frac{dr}{d\tau}
  &=0 \, , \label{eq:timeeqn3}\\
   \frac{d^2\phi}{d\tau^2}  + \left\{ \frac{a+b+2c}{r} - \frac{c(\mathcal{H} + \mathcal{F})m}{2r^2} \right\}  \frac{d\phi}{d\tau}\frac{dr}{d\tau} &=0 \, , \label{eq:phieqn3}\\
      \frac{d^2r}{d\tau^2}
    +  \left\{ \frac{2c}{r} - \frac{m\left[ c\mathcal{H} + (a+c)\mathcal{F} \right]}{2r^2} \right\} \left(\frac{dr}{d\tau}\right)^2  - \frac{b\mathcal{H}m}{2r^2}   \left(\frac{dt}{d\tau}\right)^2  +  b ( \mathcal{F}m  - r) \left(\frac{d\phi}{d\tau}\right)^2  &=0 \, . \label{eq:radialeqn3}
    \end{align}
 \end{subequations}
Thus we arrive at two constants, $E_0$ and $L_0$, of motion through the equations (\ref{eq:timeeqn3}) and (\ref{eq:phieqn3}) which represent energy and angular momentum conversations,
 \begin{subequations}
 \begin{align}
      \frac{dt}{d\tau} &\simeq \frac{E_0}{r^{2c}} \left\{1- \frac{m\left[ (a+b)\mathcal{H} + c(\mathcal{H} + \mathcal{F}) \right]}{2r} \right\}  \, , \label{eq:energycons}\\
       \frac{d\phi}{d\tau} & \simeq \frac{L_0}{r^{a+b+2c}} \left\{ 1 -\frac{mc(\mathcal{H} + \mathcal{F})}{2r} \right\} \, . \label{eq:angmomentcons}
 \end{align}
 \end{subequations}
We keep the terms only up to the first order. We now substitute the equations (\ref{eq:energycons}) and (\ref{eq:angmomentcons}) to the radial equation (\ref{eq:radialeqn3}) 
 \begin{align}
      \frac{d^2r}{d\tau^2}
    +&  \left\{ \frac{2c}{r} - \frac{m\left[ c\mathcal{H} + (a+c)\mathcal{F} \right]}{2r^2} \right\} \left(\frac{dr}{d\tau}\right)^2  - \frac{b\mathcal{H}mE_0^2}{2r^{2+4c}}    \left\{1- \frac{m\left[ (a+b)\mathcal{H} + c(\mathcal{H} + \mathcal{F}) \right]}{r} \right\} \nonumber \\ 
    &+ \frac{ b L_0^2 ( \mathcal{F}m  - r)}{r^{2(a+b+2c)}} \left\{ 1 -\frac{mc(\mathcal{H} + \mathcal{F})}{r} \right\}  =0 \label{eq:radialeqn4}
 \end{align}
Integration of this second order differential equation yields a first order differential equation
 \begin{align}
     r^{4c} &\left( 1 + \frac{m\left[ c\mathcal{H} + (a+c)\mathcal{F} \right]}{r} \right)  \left( \frac{dr}{d\tau}\right)^2 + \frac{bm \mathcal{H} E_0^2}{r} \nonumber \\
     +& \frac{2bL_0^2}{r^{2(a+b)-2}} \left\{ \frac{(1-a)m \mathcal{F}}{[1-2(a+b)]r} - \frac{1}{2(1-a-b)} \right\} -  E_0^2 + \varepsilon =0 \label{eq:radialeqn5}
 \end{align}
where $\varepsilon$ is a constant of integration. When we look at the linearized Schwarzschild limit of general relativity by setting $a=b=1$, $c=0$, $\mathcal{H}=-\mathcal{F}=-2$, we observe that the case of $\varepsilon=1$ corresponds to a time-like geodesic for massive object and $\varepsilon=0$ to a null geodesic for photon.

Furthermore, in order to gain some insights on dark matter problem we compute speed of a star at a distance $r$ from the center of gravitating source via the equations (\ref{eq:angmomentcons}) and (\ref{eq:radialeqn5}), $v^2 = (dr/d\tau)^2 + r^2 (d\phi/d\tau)^2$, as follows
 \begin{align}
     v^2 =& \frac{E_0^2 - \varepsilon}{r^{4c}} - \frac{m\{ bE_0^2 \mathcal{H} + (E_0^2 - \varepsilon)[c\mathcal{H} + (a+c)\mathcal{F}]\}}{r^{4c+1}} + \frac{(a-1)L_0^2}{r^{2(a+b+2c)-2}} \nonumber \\
     &- \frac{m L_0^2}{r^{2(a+b+2c)-1}} \left\{ c(\mathcal{F}+\mathcal{H}) + \frac{2b(1-a)\mathcal{F}}{1-2a-2b} + \frac{b[c\mathcal{H} + (a+c)\mathcal{F}]}{1-a-b} \right\}
 \end{align}
It is seen that for some values of our peculiar parameters, as the distance of a body to the source increases, its speed approaches a constant, as the observations reveal. For example, in a case of $c=0$ and $a+b \geq 1$, as $r \to \infty$, $v \to \sqrt{E_0^2 - \varepsilon}$. In a different case, $c> 0$ and $a+b+2c=1$, $\lim_{r \to \infty}v \to L_0 \sqrt{(a-1)/(a+b-1)}$.

In addition to brief analysis of dark matter we now would like to investigate the orbit equation. In the equation (\ref{eq:radialeqn5}) we write $dr/d\tau = (d\phi/d\tau)(dr/d\phi)$ and there use the equation (\ref{eq:angmomentcons}). Thus
 \begin{align}
     \left( \frac{dr}{d\phi} \right)^2 + \frac{b}{a+b-1}r^2 =& \frac{E_0^2 - \varepsilon}{L_0^2} r^{2(a+b)} - \frac{m[b \mathcal{H}E_0^2 + a \mathcal{F}(E_0^2 - \varepsilon)]}{L_0^2} r^{2(a+b)-1} \nonumber \\
     &+ \frac{b (3a+2b-2) m \mathcal{F}}{1-3(a+b)+2(a+b)^2}r \, .
 \end{align}
By changing the dependent variable as $ r= 1/u$ and rearranging terms we find
  \begin{align}
     \left( \frac{du}{d\phi} \right)^2 + \frac{b}{a+b-1}u^2 =& \frac{E_0^2 - \varepsilon}{L_0^2} u^{4-2(a+b)} - \frac{m[b \mathcal{H}E_0^2 + a \mathcal{F}(E_0^2 - \varepsilon)]}{L_0^2} u^{5-2(a+b)} \nonumber \\
     &+ \frac{b (3a+2b-2) m \mathcal{F}}{1-3(a+b)+2(a+b)^2}u^3 \, .
 \end{align}
In order to obtain an equation in the form of Newton's orbit equation we derive this with respect to $\phi$
      \begin{align}
     \frac{d^2u}{d\phi^2}  + \frac{b}{a+b-1}u =& \frac{[2-(a+b)](E_0^2 - \varepsilon)}{L_0^2} u^{3-2(a+b)} + \frac{3b (3a+2b-2) m \mathcal{F}}{2[ 1-3(a+b)+2(a+b)^2]}u^2 \nonumber \\
     & - \frac{m[5-2(a+b)][b \mathcal{H}E_0^2 + a \mathcal{F}(E_0^2 - \varepsilon)]}{2L_0^2} u^{4-2(a+b)}  \, .
 \end{align}
In this step we look at the Schwarzschild limit by setting $a=b=1, \; \mathcal{H}=-\mathcal{F}=-2$ and find the result
  \begin{align}
     \frac{d^2u}{d\phi^2}  + u = \frac{\varepsilon m}{L_0^2}  + 3 m u^2 \; .
 \end{align}
This is exactly the orbit equation of planets  for $\varepsilon=1$ and of light for $\varepsilon=0$ in the Schwarzschild metric of general relativity. Of course, one can obtain new forms of orbit equation for new settings different from above. However, we notice that as long as we stay within the Newtonian limit, which we must observe in the solar system, all possible choices yield the same form of orbit equations. By Newtonian limit we mean that the term $u^0$ must appear on the right hand side. Correspondingly, there are only two options; (i) $a+b=2$ and (ii) $a+b=3/2$. One more remark is that while our free parameter $c$ appears in our galactic velocity analysis, here it disappears automatically. Exact solutions and more detailed analysis is left for our future research projects.

\section{Concluding remarks}

In general relativity it is commonly assumed that idealised massive spinless test particles have spacetime histories that coincide with time-like geodesic (autoparallel) curves associated with the spacetime metric when acted on by gravitation alone. On the other hand, in modified theories of gravity formulated in a non-Riemannain spacetimes with torsion and/or nonmetricity, there is not a consensus on which equation, geodesic or autoparallel, describes the correct worldlines of spinless test particles under the influence of a gravitating source. In Einstein’s pseudo-Riemannian description of gravitation the affine parameter, $\tau$, physically is taken as proper time which is measured by a standard clock that is modeled by any time-like curve $x^\mu(\tau)$. Thus in spite of the fact that proper time passed between two events connected by $x^\mu(\tau)$ is path dependent, a standard Einsteinian clock accepts a proper time parameterization independent of its trajectory. On the other hand, in a non-Riemannian geometry containing nonmetricity the identification of a clock as a device for measuring proper time requires more care. In this context in a recent preprint \cite{quiros2021} it is argued that the symmetric teleparallel theories of gravity do not represent phenomenologically viable descriptions of nature due to the second clock effect. Nevertheless, in \cite{tucker1996},\cite{tucker1998} the definitions of standard clocks in theories of gravitation formulated in the non-Riemannian spacetime geometries are discussed and it seems possible to define a new standard clock that will be able to measure the affine parameter by considering the invariance of action and the full connection under gauge symmetries.  Additionally, in lemma 2 of reference \cite{koivisto2020} it is shown that in symmetric
teleparallel spacetimes, the second clock effect does not arise. Also the aurhors of \cite{adelhom2020} conclude that the definition of proper time for Weyl invariant spacetimes naturally extends to spaces with arbitrary nonmetricity. In the very recent papers \cite{hobson2020}, \cite{hobson2021} it is argued that the second clock effect does not occur in Weyl gauge theories of gravity, which are invariant both under local Poincare transformations and local changes of scale.

Thus, in this paper we discussed the bounded orbital motion of a massive spinless test particle in the background of a symmetric teleparallel geometry in the coincident gauge in terms of worldlines that are autoparallels and geodesics of nonmetric compatible spacetime connections. With a novel suggestion, we waived keeping the length of a tangent vector the same during parallel transport along the curve by the equation (\ref{eq:DTmu}) containing three free parameters, $a,b,c$. Then we obtained the autoparallel equations and geodesic equations explicitly in any coordinate chart $\{x^\mu\}$. Finally we observed that they both can match to the geodesic equation of the Riemannian geometry for some certain values of the parameters $(a=b=1, \; c=0)$ in (\ref{eq:DTmu}). We started applying our equation to a stationary, spherically axisymmetric  metric up to the first order. After a while we pass to a spherically symmetric static metric in order to avoid complexity. Finally we remarked on dark matter and orbit equation. We left the investigation of exact solutions to our autoparallel equations to our future research projects. Of course, new generation experiments with very high sensitivity will decide which curve governs correctly the motion of a spinless test particle \cite{belenchia2021}.

\section*{Acknowledgement}

The authors thank to emeritus Prof.Dr. Tekin Dereli for stimulating conversations on the non-Riemannian geometries, and the geodesic and autoparallel curves, also to the anonymous referee for instructive criticisms and to Prof.Dr. Bayram Tekin for discussion on the Schwarzschild geodesics and to Prof.Dr. Ozcan Sert for his help on usage of Maple.

\end{document}